\documentclass{iaus}
\usepackage{graphicx}
\usepackage{natbib}
\graphicspath{{./fig/}}
\usepackage{bm}
\newcommand{\BoldVec}[1]{\mathchoice%
  {\mbox{\boldmath $\displaystyle     #1$}}%
  {\mbox{\boldmath $\textstyle        #1$}}%
  {\mbox{\boldmath $\scriptstyle      #1$}}%
  {\mbox{\boldmath $\scriptscriptstyle#1$}}%
}
\newcommand{\EQ}{\begin{equation}}
\newcommand{\EN}{\end{equation}}
\newcommand{\EQA}{\begin{eqnarray}}
\newcommand{\ENA}{\end{eqnarray}}

\newcommand{\Eqs}[2]{Eqs.~(\ref{#1}) and~(\ref{#2})}

\newcommand{\Fig}[1]{Fig.~\ref{#1}}

\def\hatL{\hat{\bm{L}}}
\def\hatR{\hat{\bm{R}}}
\newcommand{\bx}{{\bm x}}
\newcommand{\bq}{{\bm q}}

\newcommand{\BB}{\BoldVec{B} {}}

\def \L    {\bm{L}}
\def \R    {\bm{R}}
\def \cH {\mathcal{H}}
\def \cHl {\mathcal{H}_L}
\def \cHr {\mathcal{H}_R}
\def \mus {\mu_{\ast}}
\def \El {E_L}
\def \Er {E_R}
\def \cL    {\mathcal{L}}

\newcommand{\ddt}[1]{\frac{d #1}{dt}}
\newcommand{\deldelt}[1]{\frac{\partial #1}{\partial t}}
\newcommand{\fder}[2]{\frac{\delta #1}{\delta #2}}
\def \curl {{\bm \nabla}\times}

\newcommand{\grav}{\BoldVec{g} {}}

\newcommand{\OO}{\BoldVec{\Omega} {}}

{}

\newcommand{\dd}{{\rm d} {}}

\def\half{{\textstyle{1\over2}}}

\newcommand{\ymn}[3]{, Mon.\ Not.\ R.\ Astron.\ Soc.\ { #2}, #3 (#1).}

\newcommand{\ypre}[3]{, Phys.\ Rev.\ E {#2}, #3 (#1).}

\newcommand{\yphl}[3]{, Phys.\ Lett.\ {#2}, #3 (#1).}

\newcommand{\yoleb}[3]{, Orig. Life Evol. Biosph. {#2}, #3 (#1).}

\newcommand{\yjour}[4]{, #2 {#3}, #4 (#1).}

\title[Symmetry breaking in the Tayler instability] %% give here short title %%
{Spontaneous chiral symmetry breaking in the Tayler instability}

\author[Del Sordo et al.]
{Fabio Del Sordo$^{1,2}$,
 Alfio Bonanno$^{3}$,
 Axel Brandenburg$^{1,2}$,
 \and Dhrubaditya Mitra$^{1}$}
\affiliation{$^1$Nordita, Roslagstullsbacken 23,
SE-10691 Stockholm, Sweden, email: {\tt fabio@nordita.org}\\
$^2$Department of Astronomy, Stockholm University,
SE 10691 Stockholm, Sweden\\
$^{3}$INAF- Catania Astrophysical Observatory, Via S.Sofia 78, 95123 Catania ITALY}

\pubyear{2011}
\volume{286}  %% insert here IAU Symposium No.
\pagerange{464--467}
\jname{Comparative Magnetic Minima}
\editors{C. H. Mandrini, eds.}
\doi{}

\begin{document}

\maketitle

\begin{abstract}
The chiral symmetry breaking properties of the Tayler instability are
discussed.
Effective amplitude equations are determined in one case.
This model has three free parameters that are determined numerically.
Comparison with chiral symmetry breaking in biochemistry is made.
\keywords{Sun: magnetic fields, dynamo, magnetic helicity}
%% add here a maximum of 10 keywords, to be taken form the file <Keywords.txt>
\end{abstract}

\firstsection % if your document starts with a section,
              % remove some space above using this command.

\section{Introduction}

An important ingredient to the solar dynamo is the $\alpha$ effect.
Mathematically speaking $\alpha$ is a pseudo scalar that can be constructed
using gravity $\grav$ (a polar vector) and angular velocity $\OO$
(an axial vector): $\grav\cdot\OO$ is thus a pseudo scalar and
is proportional to $\cos\theta$, where $\theta$ is the colatitude.
This pseudo scalar changes sign at the equator.
This explanation for large-scale astrophysical dynamos works well and therefore
one used to think that the existence of the $\alpha$ effect in dynamo theory
requires always the existence of a pseudo scalar in the problem.
This has indeed been general wisdom, although it has rarely been emphasized
in the literature.
That this is actually not the case has only recently been emphasized
and demonstrated.
One example is the magnetic buoyancy instability in the {\em absence} of
rotation, but with a horizontal magnetic field $\BB$ and vertical
gravity $\grav$ being perpendicular to each other,
so the pseudo scalar $\grav\cdot\BB$ vanishes \citep{cha+mit+bra+rhe11}.
Another example is the Tayler instability of a purely toroidal field
in a cylinder \citep{gel+rud+hol11}.
Thus, the magnetic field is again perpendicular to all possible polar
vectors that can be constructed, for example the gradient of the magnetic
energy density which points in the radial direction.
In both cases, kinetic helicity and a finite $\alpha$, both of either sign,
emerge in the nonlinear stage of the instability.
In the former case, the $\alpha$ tensor has been computed using the
test-field method.
In the latter, the components of the $\alpha$ tensor have been computed
using the imposed-field method \cite[see][for a discussion of possible
pitfalls in the nonlinear case]{HDSKB}.

The purpose of the present paper is to examine spontaneous chiral
symmetry breaking in the Tayler instability and to estimate numerically
the coefficients governing the underlying amplitude equations.
This allows us then to make contact with a model system of chemical
reactions that can give rise to the same type of spontaneous
symmetry breaking.

The connection with chemical systems is of interest because the
question of spontaneous symmetry breaking has a long history ever
since \cite{Pas53} discovered the preferential handedness of certain
organic molecules.
The preferential handedness of biomolecules 
is believed to be the result of a bifurcation event that 
took place at the origin of life itself \citep{Kon,San,BAHN}.

\section{Numerical simulations}

Our setup consists of an isothermal cylinder with a radial extent 
from $s_{\rm in}$ to $s_{\rm out}$ and vertical size $h$.
We solve the time dependent ideal MHD equations
with periodic boundary conditions in $z$, reflection in $s$ and 
periodic in $\varphi$  and a resolution ranging 
from $64^3$ to $128^3$ in the three directions.

The azimuthal field in the basic state is taken of the form
$$B_\varphi= b_0 \; (s/s_0) \exp [-(s-s_0)^2/\sigma^2]$$
with $b_0$ being a normalization constant;
the axial field $B_z$ is chosen to be zero.
In the basic state, the Lorentz force is balanced
with a gradient of pressure, and we have checked that our 
setup was numerically stable if no perturbation was introduced in the system. 
For the actual calculations we have chosen $h=2$, $s_{\rm in}=1$, $s_{\rm out}=3$, 
$s_0=2$ and $\sigma^2=0.2$. 
The sound speed is assumed to be  much larger than the Alf\'en speed 
($\approx$ ten times), similar to what happens in stellar interiors.

\begin{figure}
\begin{center}
\includegraphics[width=.6\columnwidth]{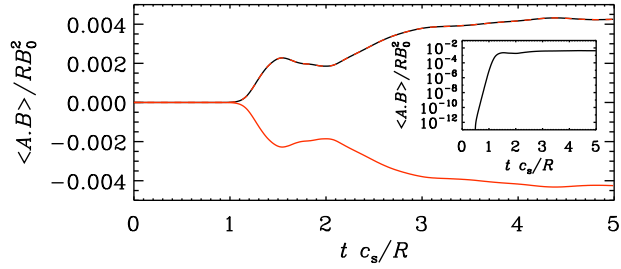}
\end{center}
\caption{
%Evolution of magnetic, current and kinetic helicities for two initial conditions
%FDS:
Evolution of magnetic helicity for two initial conditions.
differing only in the parity of their initial perturbations. 
After the exponential growth magnetic helicity levels off.
%FDS
%while current and kinetic helicities present an oscillatory decay.
In the inset a detail of the exponential growth phase.
%AB: added this
Here, $R\equiv s_{\rm in}$ is used.
}\label{fig:pcomp_axel}

\end{figure}

At the beginning of the simulation we perturb the magnetic field.
We add a perturbation of amplitude $10^{-7}$ that of the background field.
The perturbing field has a given helicity that is either positive or negative.
During the development of the instability we observe a net increase of
the helicity, as shown in \Fig{fig:pcomp_axel} where we plot
time series of the normalized magnetic helicity, which exhibits an initial exponential growth, reaches a peak and then levels off.

\section{Amplitude equations}

The linear stability analysis of this instability shows that there
exists helical growing modes.  
But the left handed and right handed
modes have exactly the same growth rate independent of their helicity. 
Hence the growth of helical perturbations cannot be described by a linear
theory.  
However a weakly nonlinear theory is able to describe it as we show below. 
Let us begin by considering two helical modes of right handed and left handed
variety respectively
each of which satisfy the Beltrami relation 
$\curl \R = \Lambda \R $ and $ \curl \bm{L} = -\Lambda \bm{L}$ .
We can deal with the Fourier transform of these modes, given by
\begin{equation}
\bm{L} (\bx) = \int \hatL(\bq) d^d q \quad \mbox{and}\quad \R (\bx) = \int \hatR(\bq) d^d q 
\end{equation}
For the left helical mode, total helicity and energy are given by 
\begin{equation}
\El =\frac{1}{2} \int \L^2 (\bx) d^d x  = \frac{1}{2} \int \hatL \cdot \hatL^{\ast} d^d q \quad\mbox{and}\quad
\cHl = \int \L \cdot \curl \L d^d x = - 2\Lambda \El,
\label{eq:eandh}
\end{equation}
where $\ast$ denotes complex conjugation. 
We then have $E = \El + \Er$ being the total energy and $\cH = \cHl + \cHr$ the total helicity.
An analogous relation holds for the right-handed helical mode too. 

In the weakly nonlinear regime the evolution of these modes  can be described by general
equations of the form: 
\begin{equation}
\deldelt{\hatL} = \fder{\cL}{\hatL} \quad\mbox{and}\quad
 \deldelt{\hatR} = \fder{\cL}{\hatR},
\label{eq:dlr} 
\end{equation}
where the Lagrangian $\cL$ can often by written down from symmetry 
considerations. 
In the present case one has to consider the fact that under parity transformation
$L$ can $R$ interchanges into each other. 
With this additional symmetry the simplest Lagrangian takes the following form
\citep{fau+dou+thu91}
\begin{equation}
\cL[\hatL,\hatR] = \int \gamma \left[|\hatL|^2 + |\hatR|^2 \right]
      -\mu \left[|\hatL|^4 + |\hatR|^4 
    -\mus |\hatL|^2|\hatR|^2 d^d q
\right]  d^d q,
\end{equation}
The coefficients $\gamma$, $\mu $ and $\mus$ cannot be found from symmetry 
considerations. 
Note that in order to show the simplest form, in writing down the Lagrangian we have 
ignored dissipation. 
This gives rise to the following set of amplitude equations,
\begin{equation}
\label{eq:evol}
\deldelt{\hatL} = \gamma \hatL -\left(\mu|\hatL|^2 + \mus|\hatR|^2\right)\hatL,
\label{eq:lt} \quad
\deldelt{\hatR} = \gamma \hatR -\left(\mu|\hatR|^2 + \mus|\hatL|^2\right)\hatR.
\label{eq:rt}
\end{equation}
For certain range of parameters these coupled equations allow the growth of one
mode at the expense of the other \citep{fau+dou+thu91}, a phenomenon known to 
biologists by the name ``mutual antagonism'' \citep{Frank}. 

Using \Eqs{eq:eandh}{eq:evol} and defining $H=\cH/{2\Lambda}$ we can obtain evolution equations for
$E$ and $H$ as
\begin{eqnarray}
\ddt{E} &=&  2\gamma E - 2(\mu + \mus)E^2 -2(\mu - \mus)H^2, \label{eq:et}\\
\ddt{H} &=& 2 \gamma H - 4\mu EH. \label{eq:ht}
\end{eqnarray}
Hence, by calculating the total energy and helicity from direct numerical 
simulations (DNS) we can determine the unknown coefficients $\gamma,\mu$ and 
$\mus$. 

\begin{figure}
\begin{center}
\includegraphics[width=.8\columnwidth]{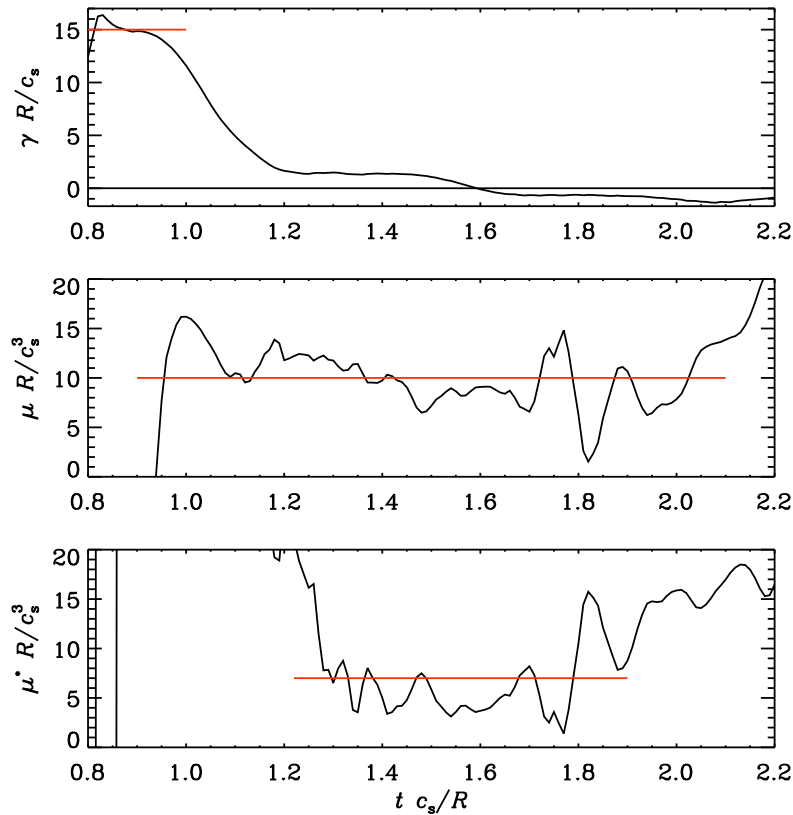}
\end{center}\caption[]{
Time dependence of $\gamma$, $\mu$, and $\mu_8$, normalized
in terms inner radius and sound speed.
The red lines give the fit results $\gamma\approx14$, $\mu\approx10$,
and $\mu^*\approx7$ in the appropriate units.
}\label{pcomp_axel3}\end{figure}

To determine the coefficients $\gamma$, $\mu$, and $\mu_*$, we define
the instantaneous logarithmic time derivatives of $E$ and $H$,
$\gamma_E=\half\dd\ln E/\dd t$ and $\gamma_H=\half\dd\ln H/\dd t$, so
we have
\EQ
\gamma=\gamma_H+2\mu E,\quad
\mu=(\gamma-\gamma_H)/2 E,\quad
\mu_*=[(\gamma-\gamma_E)E-\mu(E^2+H^2)]/(E^2-H^2).
\EN
The result is shown in \Fig{pcomp_axel3}, where we can identify
first the value of $\gamma\approx14$ during the initial linear growth
phase of the instability, and then the values $\mu\approx10$ and
$\mu_*\approx7$ during the nonlinear stage.

\section{Conclusions}
The present work has demonstrated that the Tayler instability
can produce parity-breaking and that it is possible to determine
empirical fit parameters that reproduce the nonlinear evolution of energy
and helicity.
So far, no rigorous derivation of the amplitude equations exists,
so this would be an important next step.
Comparing with the chiral symmetry breaking instabilities in biochemistry,
an important difference is that in the present equations the nonlinearity
is always cubic, while in biochemistry the dominant nonlinearity tends to
be quadratic.
In this light, it would be useful to assess more closely the possible
differences between biochemical and magnetohydrodynamical symmetry breaking.

\end{document}